\documentclass[twocolumn]{WileyMSP-template}
\usepackage{amsmath}
\usepackage{amssymb}
\usepackage{xcolor}

\usepackage{pdfpages}
\usepackage[T1]{fontenc}
\usepackage{lmodern}
\usepackage{microtype}
\begin{document}

\pagestyle{fancy}

\title{Observation of Discrete \\1D Solitons in an Optically \\Induced Lattice in Rubidium\\Atomic Vapor}

\maketitle


\author{Vjekoslav Vuli\'{c}}
\author{Neven \v{S}anti\'{c}}
\author{Hrvoje Buljan}
\author{Damir Aumiler}


\dedication{}


\begin{affiliations}
V. Vuli\'{c}, Dr. N. \v{S}anti\'{c}, prof. D. Aumiler\\
Centre for Advanced Laser Techniques\\
Institute of Physics\\
Bijeni\v{c}ka cesta 46, 10000 Zagreb, Croatia\\
Email Address: aumiler@ifs.hr

Academician H. Buljan\\
Department of Physics\\
Faculty of Science, University of Zagreb\\
Bijeni\v{c}ka cesta 32, 10000 Zagreb, Croatia\\

\end{affiliations}



\begin{abstract}
The manipulation of light in periodic structures is fundamental to the development of discrete photonics and provides a versatile platform for controlling light propagation in integrated and quantum photonic systems. 
This work reports the experimental observation of discrete one-dimensional (1D) solitons in a photonic lattice, optically induced in warm rubidium vapor. 
The lattice is generated by the interference of two coupling laser fields intersecting at a small angle, which creates a spatially modulated 1D refractive index. 
When a probe beam is focused into a single lattice site, discrete diffraction is observed. 
By increasing the probe intensity, discrete solitons emerge as a result of the balance between discrete diffraction and self-focusing within the nonlinear atomic medium. 
Experimental results are supported by numerical simulations, in which the refractive index is modeled via optical Bloch equations for a multilevel atomic system driven by the coupling and probe fields in a $\Lambda$ configuration.
These results, combined with the inherent controllability of gain and loss in atomic vapors, suggest that this platform provides a versatile foundation for exploring non-Hermitian nonlinear dynamics in photonic lattices.
\end{abstract}

\section{Introduction}
In the last 30 years, photonic lattices have emerged as a versatile platform for designing and controlling the behavior of light in frequency and real space \cite{christodoulides2003, lederer2008, longhi2009, garanovich2012}.
These periodic dielectric structures thus hold great potential for development of new optical devices and accelerating advancements in emerging areas such as topological photonics \cite{ozawa2019}. 
The behavior of light in a photonic lattice is mathematically analogous to the motion of electrons in a crystal. 
Consequently, these structures serve as a versatile laboratory tool for investigating classical optical analogues of quantum phenomena originally encountered in atomic and condensed matter physics \cite{christodoulides2003, lederer2008, longhi2009, garanovich2012}.
Such phenomena include Anderson localization \cite{segev2013}, Bloch oscillations \cite{morandotti1999}, and Zener tunneling \cite{trompeter2006}, as well as more recent explorations into non-Hermitian dynamics within parity-time-symmetric systems \cite{elganainy2018}.

Various experimental approaches facilitate the realization of one- and two-dimensional (1D and 2D) photonic lattice structures. 
These include waveguide arrays etched in semiconductors \cite{millar1997}, optically induced lattices in photorefractive crystals  \cite{fleischer2003}, titanium in-diffusion in lithium niobate crystals \cite{iwanow2004}, femtosecond laser-written arrays in silica \cite{pertsch2004, szameit2005}, and arrays of optical fibers \cite{pertsch2004prl}. 
Optically induced lattices in photorefractive crystals are of significant interest due to their capacity for creating dynamic, reconfigurable waveguide arrays \cite{fleischer2003}.
In these systems, the lattice geometry is directly governed by the interference pattern of the inducing laser beams, offering a level of flexibility not easily achieved in static fabricated structures.

An analogous approach in atomic media is increasingly used in which optical induction relies on the internal atomic structure and the accompanying coherent phenomena such as electromagnetically-induced transparency (EIT) \cite{sheng2015}. 
In this approach, the EIT effect is used to spatially modify the refractive index of Doppler-broadened rubidium atoms in a vapor cell. 
Due to a broad range of available tunable parameters (atom concentration, laser intensities and detunings, and beam geometry), multi-level atomic systems provide a versatile tool for studying optically induced photonic lattices. 
Moreover, the ability to produce gain \cite{zhang2016} fundamentally distinguishes atomic systems from alternative photonic lattice platforms, establishing them as a unique platform to investigate a plethora of phenomena in light-matter interactions and condensed matter physics, as well as a platform for designing reconfigurable photonic devices and the development of novel photonic devices targeting an active light control \cite{suchkov2016}.

To date, various phenomena in atomic systems have been studied theoretically, such as optical Bloch oscillation and Zener tunneling \cite{zhang2017}, and honeycomb lattices \cite{zhang2015}.
Various lattice configurations have also been demonstrated experimentally, including 1D lattices \cite{sheng2015}, square \cite{ning2021}, honeycomb \cite{feng2023}, Kagome \cite{liang2023b}, and Lieb \cite{liang2023} geometries. 
These implementations enabled the observation of diverse phenomena, such as 1D and 2D discrete diffraction \cite{sheng2015,yuan2019}, the electromagnetically induced Talbot effect \cite{zhang2018}, optical Bloch oscillations \cite{zhang2022}, non-Hermitian twisted photonic lattices \cite{yuan2026}, and parity-time symmetric lattices \cite{zhang2016}.
More recently, edge solitons were observed in a photonic graphene lattice where Raman gain was introduced to compensate for the significant absorption experienced by the edge state during propagation \cite{zhang2020}. 

Discrete spatial solitons were previously observed and studied in solid-state waveguide arrays \cite{eisenberg1998, morandotti1999} and subsequently in photorefractive crystals \cite{fleischer2003}. 
In contrast, the investigation of discrete solitons in atomic systems remains largely unexplored. 
The unique tunability of atomic systems offers opportunities for investigating soliton phenomena involving resonant light–matter interactions, including dissipative discrete solitons and $\mathcal{PT}$-symmetric discrete solitons where gain and loss play a crucial role.

This work presents the experimental observation of discrete diffraction and one-dimensional (1D) discrete solitons in an optically induced lattice within rubidium atomic vapor.
The lattice is formed through the interference of two coupling laser beams intersecting at a small angle inside the vapor cell, while the transverse intensity profile of a probe beam is monitored during its propagation through the induced structure, see Figure \ref{fig:scheme}.
To reduce probe absorption a far-detuned EIT $\Lambda$ scheme is employed, with both the coupling and probe fields detuned by approximately 1 GHz from the single-photon resonance (i.e., in the far wings of the Doppler-broadened transition).
When the lattice is absent, the probe beam, focused at the cell entrance, undergoes significant diffraction.
Upon induction of the optical lattice, discrete diffraction is observed, and quantitative agreement is shown between the measured and simulated probe beam dynamics.
Finally, self-focusing of the probe beam, driven by saturable nonlinearity, is demonstrated as the probe intensity increases, culminating in the formation of a 1D soliton when self-focusing counterbalances discrete diffraction.

The structure of this paper is as follows. 
First, a theoretical model is introduced to describe the optical lattice induced by the coupling fields and experienced by the probe field.
Optical Bloch equations are employed for realistic multilevel $^{87}$Rb atoms, including the hyperfine structure.
A spatially varying coupling field leads to a spatially modulated complex refractive index encountered by the probe field. 
This calculated refractive index is subsequently integrated into the paraxial wave equation to simulate probe laser propagation within the optically induced lattice.
Quantitative agreement is shown between the measured and simulated probe beam dynamics, highlighting discrete diffraction and the formation of 1D solitons driven by saturable nonlinearity as the probe intensity increases.

\section{Theoretical description of the optical lattice}

\begin{figure}
	\includegraphics[width=1\linewidth,trim={0.3cm 0cm 0.05cm 0cm},clip]{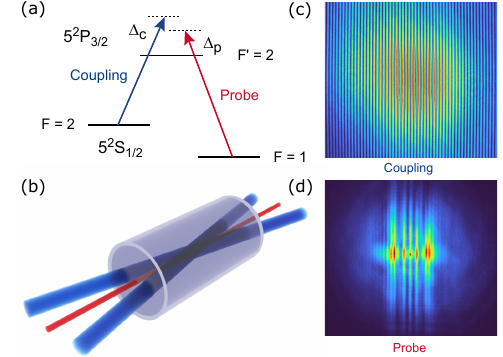}
	\caption{Experimental scheme relevant for realizing the 1D optically induced lattice in rubidium atomic vapor. Far-detuned $\Lambda$-type excitation scheme is utilized (a) to enable refractive index change felt by the probe laser by varying the coupling laser intensity. Spatially periodic coupling laser intensity (c) is obtained through the interference of two coupling laser beams that intersect under a small angle inside the glass cell containing rubidium atomic vapor (b). Probe is focused to a single lattice site at the beginning of the cell and imaged at different propagation distances. Measured probe intensity distribution after passing through the cell is illustrated in (d).}
	\label{fig:scheme}
\end{figure}

The interference of two coupling laser beams, intersecting at a small angle $\theta$ in the $xz$ plane, where $z$ is the propagation direction, generates a periodic intensity distribution along the $x$-axis, see Figure \ref{fig:scheme}(c), with a period $d=\lambda/\sin(\theta)$.
The optical response of the rubidium atoms varies according to the local intensity of the coupling field. 
Consequently, a probe beam propagating through the atomic vapor encounters a spatially periodic refractive index, i.e., an optical lattice, which changes probe transverse intensity profile as it propagates through the vapor.
To model the response of the rubidium vapor under simultaneous coupling and probe excitation, the interaction between the atomic system and the electromagnetic fields is described using the optical Bloch equations. 
This approach relates the microscopic atomic dynamics to the macroscopic susceptibility, providing a framework for calculating the complex refractive index of the vapor.

A three-level $\Lambda$-type system of $^{87}$Rb atoms is considered, which includes the ground-state hyperfine levels ($5S_{1/2}, F=1, 2$) and the excited-state hyperfine level ($5P_{3/2}, F'=2$), as shown in Figure \ref{fig:scheme}(a). 
This configuration represents a standard $\Lambda$-scheme, frequently utilized in studies of electromagnetically induced transparency (EIT) \cite{fleischauer2005}.

The optical Bloch equations of the system are given by:
\begin{equation}
\begin{split}
    \frac{d\rho_{nm}}{dt} &=\frac{-i}{\hbar}\:[\widehat H,\widehat\rho\:]_{nm}-\gamma_{nm}\rho_{nm},\qquad n\neq m, \\
	\frac{d\rho_{nn}}{dt} &=\frac{-i}{\hbar}\:[\widehat H,\widehat\rho\:]_{nn}+\sum\limits_{E_m>E_n}\Gamma_{nm}\rho_{mm}\\
                          &-\sum\limits_{E_m<E_n}\Gamma_{mn}\rho_{nn} - \Gamma_t\rho_{nn} + \Gamma_{nn} ,
	\label{Eq:obe}
\end{split}
\end{equation}

where $\rho_{nn}$ are hyperfine level populations and $\rho_{nm}=\rho_{mn}^*$ are coherences induced in the system.
The Hamiltonian of the system is $\widehat H=\widehat H_0+\widehat H_{int}$, where $\widehat H_0$ is the Hamiltonian of the free atom, and $(\widehat H_{int})_{nm}=-\mu_{nm}E(t)$ represents the interaction of the atom with probe and coupling laser electric fields $E_{probe}(t)=E_pe^{i\omega_pt}$ and $E_{coupling}(t)=E_ce^{i\omega_ct}$, with $\mu_{nm}$ the transition dipole moment.
Coherence damping constants are given by $\gamma_{nm}=(\Gamma_n+\Gamma_m)/2+\Gamma_t $, where $\Gamma_n$ and $\Gamma_m$ are total relaxation rates of levels $n$ and $m$, whereas $\Gamma_{nm}$ is population decay from level $m$ to level $n$.

Atoms ballistically traverse the coupling laser interference pattern, passing through alternating bright and dark fringes. 
Given the small lattice period ($d < 50\mu$m), the interaction time is significantly limited and transit time broadening needs to be taken into account. 
The broadening rate $\Gamma_{\text{t}} = 1/\tau_{\text{t}}$ is determined by the transit time $\tau_{\text{t}}$ of atoms through a single lattice site, defined as the ratio of the bright-fringe full-width at half-maximum (approximately $d/2$) to the mean one-dimensional atomic velocity (approximately 150 m/s at 100 °C).
For a 37 $\mu$m lattice at a vapor temperature of 100 °C, $\Gamma_{\text{t}}$ is approximately $2\pi \times 1.3$ MHz.

As atoms spontaneously decay during their transit through the dark fringes, only the ground-state populations are replenished upon re-entry into the interaction zones ($\Gamma_{nn} = 0$ for the excited state). 
Furthermore, to account for longitudinal atomic motion in the Doppler-broadened vapor, Equation \ref{Eq:obe} is solved for discrete velocity groups and subsequently averaged over the complete longitudinal Maxwell–Boltzmann velocity distribution.

The system of Equations \ref{Eq:obe} defines the time dynamics of the three-level atoms under dual-field excitation, enabling the derivation of the complex refractive index experienced by the probe beam within the induced optical lattice. 
While a steady-state analytical solution is obtainable, the resulting expressions are algebraically cumbersome and do not provide much physical insight.

Furthermore, analytical solutions in the weak-probe approximation ($E_p \ll E_c$) that can be found in the literature (see e.g. \cite{fleischauer2005}) are inapplicable to the present experimental conditions.
At the cell entrance, the probe laser has powers reaching several milliwatts and is focused into a single lattice site, resulting in regimes where $E_p > E_c$ at the cell entrance or even $E_p \gg E_c$ for higher probe powers.
In contrast, due to probe diffraction during propagation, the condition $E_p < E_c$ typically holds at the cell exit. 
To accurately capture the behavior of the system in these different regimes and the associated saturable nonlinearity, a numerical solution of Equation \ref{Eq:obe} is required.

Of particular interest is the steady state solution of Equations \ref{Eq:obe} for the coherence $\rho_{31}=\sigma_{31}e^{-i\omega_pt}$, i.e., the coherence induced by the probe laser field between the ground state $|5S_{1/2}, F=1\rangle$ and the excited state $|5P_{3/2}, F'=2\rangle$.
Here, $\sigma_{31}$ is the slowly varying envelope of the coherence, which defines the response of the atomic system to the probe field.
The macroscopic polarization of the medium induced by the probe laser is given by $N\mu_{13}\rho_{31}$, where $N$ denotes the atomic number density.
The atomic susceptibility to the probe electric field is then given by:
\begin{equation}
	\chi = \frac{N\mu_{13}}{\epsilon_0 E_p}\sigma_{31}.
\end{equation}
In the limit of a dilute medium, the complex refractive index $n$ is approximated as $n=\sqrt{1+\chi}\approx 1+\chi/2$. This leads to the expression $n=1+n_{\text{Re}}+in_{\text{Im}}$, where the real and imaginary components are given by:
\begin{equation}
	n_{\text{Re}} = \frac{N\mu_{13}}{2\epsilon_0 E_p}\text{Re}(\sigma_{31}), \quad n_{\text{Im}} = \frac{N\mu_{13}}{2\epsilon_0 E_p}\text{Im}(\sigma_{31}).
\end{equation}

The 3-level $\Lambda$ scheme (Figure \ref{fig:scheme}(a)) is a widely used approximation for describing $^{87}$Rb atoms interacting with coupling and probe laser fields. 
In this approximation the influence of excited-state hyperfine levels $5P_{3/2}, F'=0, 1,$ and $3$ is neglected. 
Under the present experimental conditions, however, this simplified model, while qualitatively predicting refractive index variations, is insufficient to reproduce experimental results. 
A full 6-level model is therefore required, in which the coupling laser field excites $5S_{1/2}, F=2 \rightarrow 5P_{3/2}, F'=1,2,3$ transitions, while the probe field excites $5S_{1/2}, F=1 \rightarrow 5P_{3/2}, F'=0,1,2$ transitions. 
This six-level model is employed for all refractive index calculations throughout this study.

\begin{figure}[ht]
	\centering
	\includegraphics[width=0.95\linewidth,trim={2.cm 3cm 14cm 1.5cm},clip]{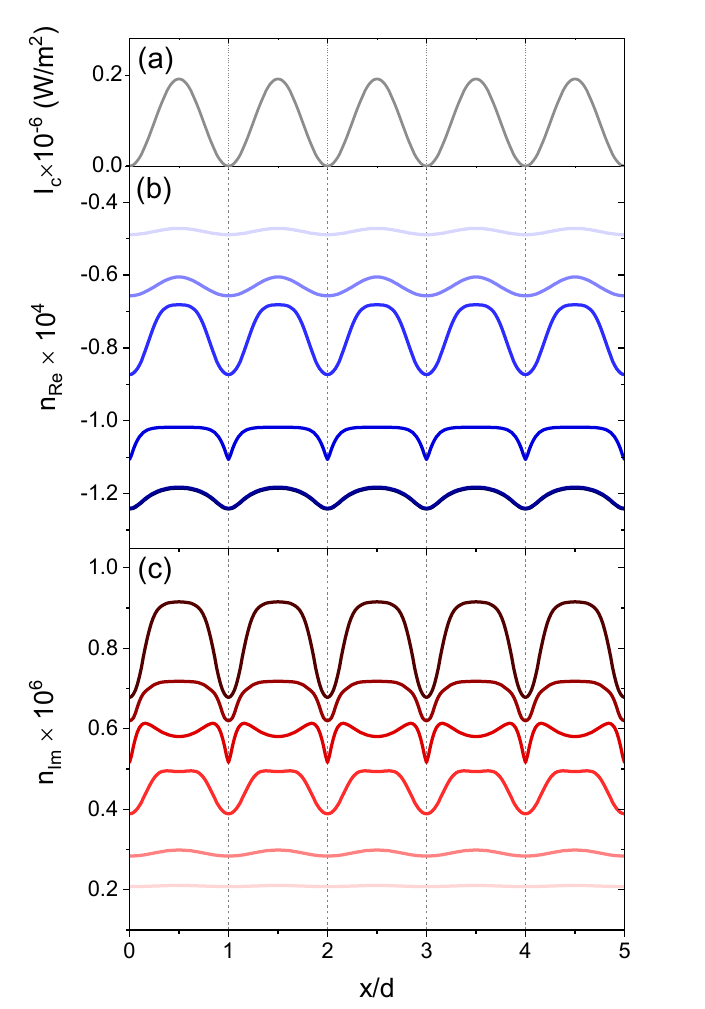}
	\caption{Optically induced lattice in rubidium atomic vapor within the multi-level $\Lambda$-type excitation scheme illustrated in Figure \ref{fig:scheme}. (a) Spatially periodic coupling laser intensity distribution. (b, c) Calculated (b) real and (c) imaginary components of the induced refractive index experienced by a spatially uniform probe laser. Curves are shown for probe field amplitudes $E_p = (0.1, 0.5, 5, 10, 15, 20) \times 10^3$ V/m, where darker colors indicate lower $E_p$ values. Other parameters: $\Delta_c$=1000 MHz, $\Delta_p$=960 MHz, $\Gamma_t=2\pi\times$1.3 MHz, $T=100$ °C.}
	\label{fig:lattice}
\end{figure}

The total refractive index incorporates contributions from all three allowed hyperfine transitions: 
\begin{equation}
	\begin{aligned}
		n_{\text{Re}}\! &=\! \frac{N}{2\epsilon_0 E_p} \operatorname{Re}\left( \mu_{13}\sigma_{31} + \mu_{14}\sigma_{41} + \mu_{15}\sigma_{51} \right), \\
		n_{\text{Im}}\! &=\! \frac{N}{2\epsilon_0 E_p} \operatorname{Im}\left( \mu_{13}\sigma_{31} + \mu_{14}\sigma_{41} + \mu_{15}\sigma_{51} \right),
	\end{aligned}
\end{equation}
where $\sigma_{31}$, $\sigma_{41}$, and $\sigma_{51}$ refer to the slowly varying coherences between the ground $5S_{1/2}, F=1$ level and excited $5P_{3/2}, F'=0,1,2$ levels, respectively.  
The calculated spatial variation of the refractive index that is induced by the coupling laser beams is shown in Figure \ref{fig:lattice}. 

In the linear regime, specifically for probe field amplitudes $E_p < 100$ V/m, both $n_{\text{Re}}$ and $n_{\text{Im}}$ remain nearly constant regardless of further reductions in $E_p$. 
Nonlinear behavior is notably more pronounced in the imaginary component $n_{\text{Im}}$, where absorption is significantly diminished at $E_p = 500$ V/m compared to $E_p = 100$ V/m (as indicated by the two darkest curves in Figure \ref{fig:lattice}(c)), whereas the corresponding curves for the real part $n_{\text{Re}}$ overlap at these amplitudes (Figure \ref{fig:lattice}(b)). 
As $E_p$ continues to increase, both the real and imaginary components of the refractive index approach zero, eventually saturating at high probe intensities. 
For the red two-photon detuning $\delta = \Delta_p - \Delta_c = -40$ MHz depicted in Figure \ref{fig:lattice}, both $n_{\text{Re}}$ and $n_{\text{Im}}$ scale positively with the coupling laser intensity. 
Consequently, the local maxima of the coupling intensity $I_c$ correspond directly to the maxima of the induced refractive index, defining the guiding sites of the photonic lattice.

For blue two-photon detunings ($\delta > 0$), $n_{\text{Im}}$ exhibits qualitatively similar behavior, but with periodic oscillations that are more pronounced and regular. 
This enhanced contrast is a direct consequence of a reduction in single photon absorption due to larger detuning, and absorption can primarily be attributed to the two-photon Raman resonance.
Simultaneously, $n_{\text{Re}}$ displays qualitatively different behavior, decreasing as the coupling laser intensity increases (see Figure S2(b) in the Supporting Information). 
This contrasting behavior of $n_{\text{Re}}$ between red and blue detunings leads to different discrete diffraction patterns as the probe beam propagates through the induced optical lattice.

\begin{figure*}[t]
	\centering
	\includegraphics[width=1.\linewidth,trim={0.cm 9.8cm 0.5cm 6.5cm},clip]{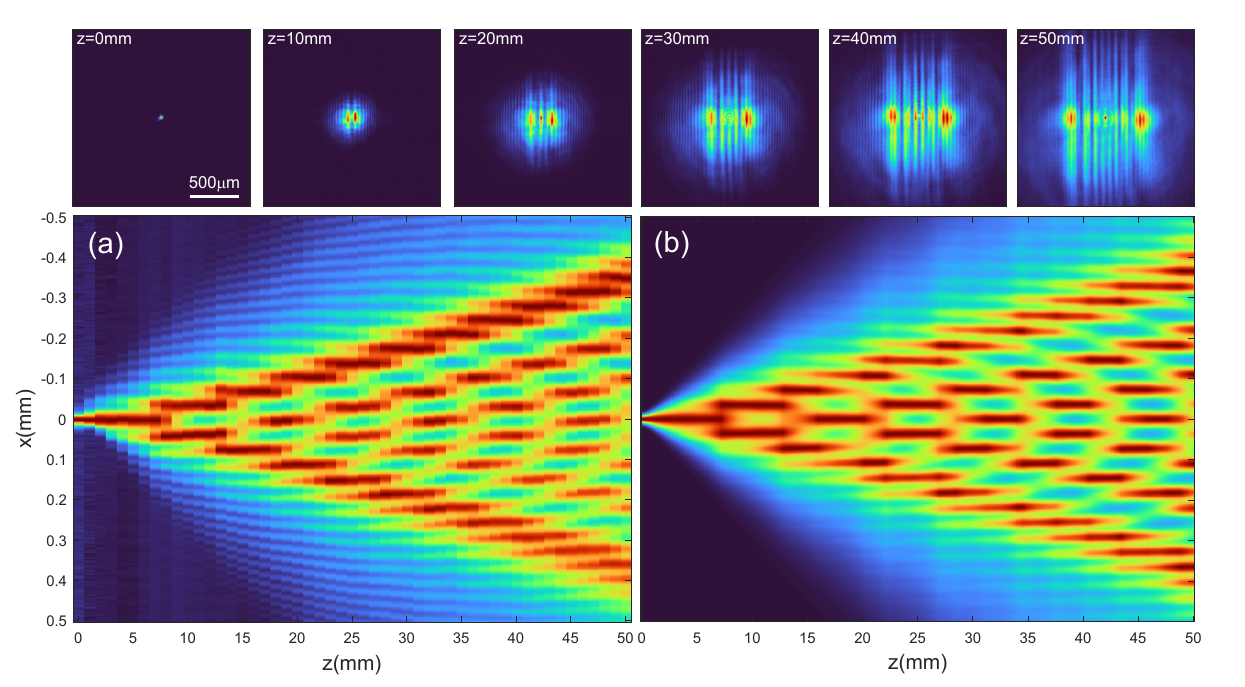}
	\caption{Discrete diffraction in an optically induced lattice in rubidium atomic vapor. Images of the probe laser beam (transversal beam intensity) at the cell entrance ($z$=0) and at different positions within the cell (propagation distances $z$=10, 20, 30, 40, 50 mm) are shown in the upper panels. Measured (a) and simulated (b) 1D probe intensity distributions are obtained by integrating over the $y$ direction, and are normalized for clarity (see Figure S4 in Supporting Information for the non-normalized discrete diffraction measurement).}
	\label{fig:diffraction}
\end{figure*}

To simulate the propagation of the probe laser within the optically induced lattice, the paraxial wave equation for the probe laser electric field  $E_{probe}(x,y,z)=E_p(x,y,z)e^{ikz}$ is employed:
\begin{equation}
	\frac{\partial E_p}{\partial z} = \frac{i}{2k}\left( \frac{\partial^2}{\partial x^2} + \frac{\partial^2}{\partial y^2} \right)E_p + ik(n_{\text{Re}} + in_{\text{Im}})E_p,
	\label{Eq:paraxial}
\end{equation}
where $k$ is the wavevector in vacuum. 
The numerical simulation is implemented in two stages. 
First, the optical Bloch equations (Equation \ref{Eq:obe}) are solved parametrically across a range of probe and coupling field amplitudes, $E_p$ and $E_c$, to establish a comprehensive map of the atomic response.
Subsequently, Equation \ref{Eq:paraxial} is solved using the split-step Fourier method \cite{agrawal}. 
In each longitudinal propagation step, the local complex refractive index values, $n_{\text{Re}}(x,y)$ and $n_{\text{Im}}(x,y)$, are evaluated via interpolation based on the local field distributions $E_c(x,y)$ and $E_p(x,y)$ and the results of Equation \ref{Eq:obe}. 
Because $n_{\text{Re}}$ and $n_{\text{Im}}$ exhibit a nonlinear dependence on the probe amplitude (as shown in Figure \ref{fig:lattice}), Equation \ref{Eq:paraxial} functions as a nonlinear wave equation, capturing the effects of saturable nonlinearity on the beam dynamics.
Attenuation of the coupling fields with $z$ as they propagate through the vapor is neglected, as no significant attenuation of the coupling beams is observed in the experiment.  

\section{Experimental results}
Experimental results of the probe laser propagation within the induced 1D optical lattice are presented in Figure \ref{fig:diffraction}. 
The upper panels display the measured transverse probe beam intensity at various propagation distances within the vapor cell. 
At the cell entrance ($z = 0$), the probe beam is focused into a dark interference fringe, subsequently exhibiting characteristic discrete diffraction as it propagates through the medium. 
Figure \ref{fig:diffraction}(a) depicts the measured 1D intensity distribution of the probe beam, obtained by integrating the transverse intensity profile along the $y$-axis, as a function of the propagation distance $z$.
For enhanced visual clarity, the 1D intensity distributions are normalized. 
The corresponding numerical simulation of the normalized 1D probe intensity distribution is shown in Figure \ref{fig:diffraction}(b). 
Note that in the absence of the optically induced lattice (coupling beams inactive), a Gaussian beam with a full-width at half-maximum (FWHM) of approximately 1 mm is observed at $z = 50$~mm. 
These measurements are performed using a coupling beam power $P_c = 200$ mW with a FWHM diameter of 1.7 mm. 
Other experimental parameters include detunings $\Delta_c = 1000$ MHz and $\Delta_p = 1070$ MHz, a vapor temperature $T = 100$~°C, and a probe power $P_p = 0.05$~mW.

The discrete diffraction pattern observed in Figure \ref{fig:diffraction}(a) is analogous to the dynamics reported in coupled waveguide arrays \cite{pertsch2002}, where discrete diffraction is analyzed using coupled-mode theory (or the tight-binding approximation) \cite{lederer2008}.

Notably, a qualitatively different discrete diffraction pattern is observed when the probe beam is focused into a bright interference fringe at the cell entrance (under otherwise identical experimental conditions), as illustrated in Figure S3 of the Supporting Information. 
In general the discrete diffraction of a probe beam blue-detuned from the two-photon resonance focused into a dark fringe is equivalent to that of a red-detuned probe focused into a bright fringe.
Similarly, the propagation of a blue-detuned probe focused into a bright fringe exhibits the same evolution as a red-detuned probe focused into a dark fringe. 
Ultimately, two distinct discrete diffraction patterns are observed: the one shown in Figure \ref{fig:diffraction} ($\delta>0$, probe focused to a dark fringe; or $\delta<0$, probe focused to a bright fringe), and the one shown in Figure S3 in Supporting Information ($\delta>0$, probe focused to a bright fringe; or $\delta<0$, probe focused to a dark fringe). 

\begin{figure}[t]
	\centering
	\includegraphics[width=1\linewidth,trim={2.cm 4.0cm 4.cm 3.0cm},clip]{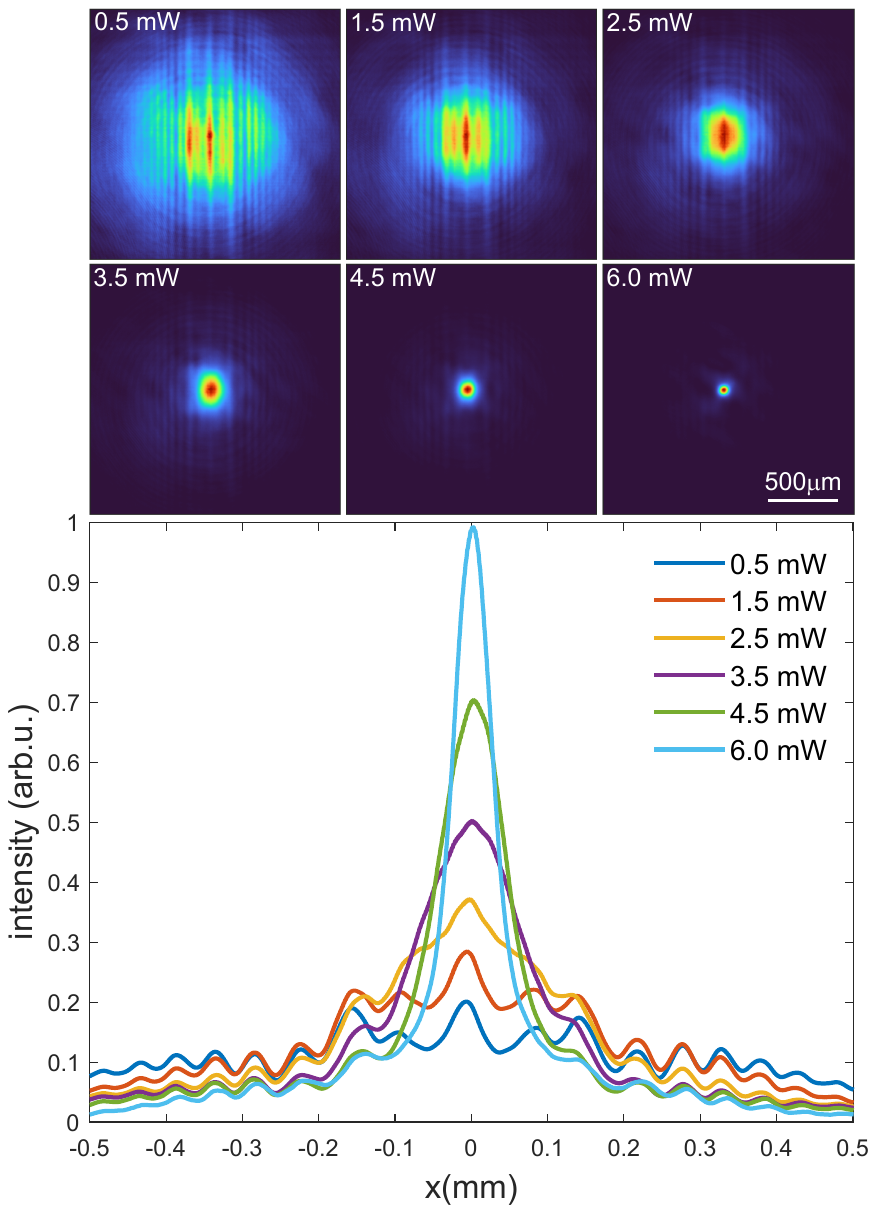}
	\caption{Self-focusing of the probe beam in an optically induced lattice in rubidium atomic vapor. Images of the probe laser beam (transversal beam intensity) after passing through the cell ($z$=50 mm) for different probe beam power ($P_p$=0.5, 1.5, 2.5, 3.5, 4.5, 6 mW) are shown in the top panels. Measured 1D probe intensity distributions are obtained by integrating over the $y$ direction and are shown in the bottom panel (see Figure S6 in Supporting Information for a numerical simulation of the nonlinear probe beam propagation dynamics).}
	\label{fig:soliton}
\end{figure}

Discrete diffraction is strongly influenced by the temperature of the cell and detuning of the coupling beams. 
The refractive index is proportional to the atom concentration, which is determined by the cell temperature. 
Heating of the cell is therefore needed in order to obtain a sufficiently high atom concentration that can result in noticeable refractive index changes. 
Refractive index changes induced by the coupling beams depend on coupling beam detuning, and are generally larger for smaller detuning $\Delta_c$. 
But smaller detuning leads to higher absorption, which effectively reduces the coupling beam intensity and consequently the induced refractive index changes. 
For a cell of 5 cm used in the experiment, optimal conditions are obtained for cell temperatures around 100 °C and 1000 MHz detuning. 
Further increasing the temperature raises the atom concentration but lowers the coupling beam intensity, so optimal discrete diffraction conditions for $T>100$~°C are obtained for $\Delta_c>1000$ MHz.

While discrete diffraction characterizes the linear regime at low probe intensities, as the probe power is increased ($P_p \gtrsim 1$ mW) the probe beam at the exit of the cell starts to change shape. 
This transition to the nonlinear regime is a direct consequence of the nonlinear change of the refractive index with $E_p$, as established in the multi-level atomic model (Figure \ref{fig:lattice}).

Under conditions of red two-photon detuning ($\delta < 0$), the probe beam undergoes pronounced self-focusing with increasing power, as illustrated in Figure \ref{fig:soliton}. 
A discrete soliton is fully formed once the probe power reaches $P_p \gtrsim 6$ mW. 
These measurements are obtained with a coupling power $P_c = 200$ mW and detunings of $\Delta_c = 1000$ MHz and $\Delta_p = 960$ MHz at a vapor temperature of 100 °C.

Notably, these nonlinear effects are observed at the milliwatt power scale, an efficiency facilitated by the nonlinearity generated in the proximity of the $5S_{1/2}, F=1 \rightarrow 5P_{3/2}, F'=0, 1, 2$ one-photon transitions. 
The magnitude of this nonlinearity is comparable to those observed in optically induced lattices within photorefractive crystals \cite{fleischer2003prl}.
In contrast to the red-detuned regime, no significant self-focusing is observed in the blue wing of the two-photon resonance, probably due to the increased detuning from the one-photon resonance, which results in a significantly reduced effective nonlinear refractive index $n_2$. 

At $P_p \approx 6$ mW, the probe beam propagates through the vapor cell with a nearly invariant spatial profile, a consequence of the balance between nonlinear self-focusing and discrete diffraction.
An initial estimate for the nonlinear refractive index could be derived from the critical power for self-trapping $P_{cr} = \pi (0.61)^2 \lambda^2 / (8 n_0 n_2)$. 
Using the experimental threshold of 6 mW, this formula yields $n_2 \approx 1.5 \times 10^{-11}$ m$^2$/W. 
The above expression, however, assumes a uniform medium and should therefore be used with caution when estimating the nonlinear refractive index in a lossy 1D optical lattice. 
Furthermore, the classical critical power formula assumes a pure Kerr nonlinearity ($n_{\text{Re}} = n_0 + n_2 I_p$), whereas in resonant atomic systems, a saturable nonlinearity provides a more accurate physical description:
\begin{equation}
	n_{\text{Re}} = n_0 + \frac{n_2 I_p}{1 + I_p / I_s},
	\label{Eq:saturation}
\end{equation}
where $I_s$ denotes the saturation intensity.
This is supported by the calculations of the refractive index $n_{\text{Re}}$ as a function of probe intensity using Equation \ref{Eq:obe}, which results in $n_2\approx1\times 10^{-10}$ m$^2$/W and $I_s\approx1.3\times 10^{6}$ W/m$^2$ (see Figure S7 in Supporting Information). 
The obtained value for $n_2$ agrees with a recent measurement of nonlinear refractive index of warm rubidium vapor \cite{kardum2026}. 

\section{Conclusion}
In summary, this work demonstrates discrete diffraction and spatial soliton formation within a reconfigurable optical lattice induced in a warm $^{87}$Rb atomic vapor. 
By selectively exciting a single lattice site and imaging the longitudinal evolution of the probe beam, we have experimentally characterized discrete dynamics that are fundamentally equivalent to those observed in coupled waveguide arrays and photorefractive crystals.

Additionally, it was shown that the system exhibits strong nonlinear propagation effects and eventually discrete soliton formation at milliwatt power scales when the probe is tuned to the red wing of the two-photon resonance. 
The theoretical framework, coupling the paraxial wave equation with a realistic multi-level atom model, accurately reproduces the measured beam dynamics. 
This high level of agreement validates the necessity of accounting for the full hyperfine structure of the $^{87}$Rb $D_2$ line and establishes warm atomic vapors as a promising and highly tunable platform for studying nonlinear lattice physics.

Beyond traditional discrete optics, this system offers unique advantages. 
Unlike solid-state waveguides and photorefractive crystals, atomic vapors allow for the incorporation of gain through complex coherent excitation schemes. 
This capability distinguishes the present platform and provides a clear pathway for investigating nonlinear discrete diffraction in non-Hermitian and $PT$-symmetric regimes.

The study of warm atomic vapors is currently experiencing a striking revival, driven by advancements in vapor-cell atomic clocks, magnetometers, and chip-scale atomic devices using microfabricated vapor cells \cite{kitching2018}. 
The results presented in this work should be viewed in the context of these recent advancements, particularly in the context of the use of warm atomic vapors for non-linear optics, including squeezed light generation \cite{ries2003}, quantum memories \cite{phillips2001}, and the study of quantum fluids of light \cite{santic2018}.
In all these applications warm atomic vapors are a media of choice as they allow for precise control of the linear and nonlinear indices of refraction through optical manipulation of the atomic states \cite{glorieux2023}.

\section{Experimental Section}

\begin{figure}[t]
	\centering
	\includegraphics[width=1\linewidth,trim={0.cm 0.0cm 0.cm 0.0cm},clip]{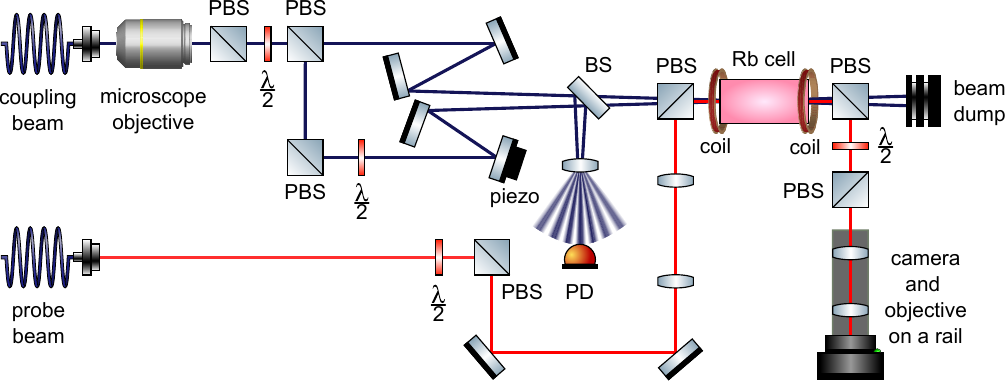}
	\caption{Experimental setup. The vapor cell filled with $^{87}$Rb is heated to 100°C. The coupling beams intersect at the center of the cell at a small angle, forming an interference pattern, part of which is expanded by a lens and sent to a photodiode for stabilization. The probe beam co-propagates with the coupling beams through the cell and is imaged by a CMOS camera. The camera and its objective are mounted on a precision rail so different planes in the propagation axis can be imaged. PBS: polarization beam splitter, $\lambda$/2: half-wave plate, piezo: piezoelectric translator, BS: beam splitter, PD: photodiode, coil: magnetic coil.}
	\label{fig:setup}
\end{figure}

The experimental setup is shown in Figure \ref{fig:setup}. 
The coupling laser (Sirah Matisse CR) and the probe laser (Toptica DL100) are both stabilized by a software PID control loop to a HighFinesse/Ångstrom WLM-VIS wavemeter to a precision of 10 MHz. 
Both laser beams are coupled into polarization maintaining fibers for mode cleanup. 
The coupling beam is collimated with a microscope objective to obtain a large diameter beam free from spherical aberrations. 
The beam is split in two identical beams with horizontal polarization and sent to intersect at a small angle (typically 2$\theta\approx0.4$°) in the middle of a vapor cell filled with \textsuperscript{87}Rb and heated to 100°C, thus creating an optically induced lattice along the transverse direction $x$. 

Part of the coupling beams is split before the vapor cell and made to interfere in a lens which expands the interference pattern so that a bright interference fringe is wider than a photodiode active area. 
The signal from the photodiode is used for side-of-fringe locking via a Newport LB1005-S servo controller, whose output is fed to a piezoelectric translator mounted on a mirror in one of the coupling beam's path. 
In this way the optically induced lattice is stabilized to movements no larger than 3 $\mu$m.

The vertically polarized probe beam co-propagates with the coupling beams through the vapor cell and is imaged with a CMOS camera (IDS UI-3240CP-NIR), with the camera and its objective mounted on a precision rail to allow for imaging of different planes along the propagation axis inside the vapor cell. 
The probe beam is focused to a FWHM waist diameter of $\approx$10 $\mu$m at the entrance of the cell. 
The coupling beams are filtered out before the camera by a PBS, half-wave plate and an another PBS. 

In order to minimize the effect of polarization rotation in the warm rubidium vapor, magnetic coils are placed at the ends of the vapor cell, with the currents through them adjusted to minimize the "leakage" of the coupling beams to the camera.
Every measurement of the transmitted probe beam is followed by a measurement with the probe off in order to eliminate stray light. 
The probe is turned on and off by an acousto-optic modulator (AOM). 
The AOM and camera triggering are both controlled by a Labjack T4 DAQ device interfaced by a Python script which also adjusts the setpoint of the wavelength lock to the wavemeter for the probe laser.

\medskip
\textbf{Supporting Information} \par 
Supporting Information is available from the Wiley Online Library or from the author.

\medskip
\textbf{Acknowledgements} \par 
N.Š., H.B., and D.A. acknowledge support from the project "Implementation of cutting-edge research and its application as part of the Scientific Center of Excellence for Quantum and Complex Systems, and Representations of Lie Algebras", Grant No. PK.1.1.10.0004, co-financed by the European Union through the European Regional Development Fund - Competitiveness and Cohesion Programme 2021-2027.
We also acknowledge support from the project "Centre for Advanced Laser Techniques (CALT)", cofunded by the European Union through the European Regional Development Fund under the Competitiveness and Cohesion Operational Programme (Grant No. KK.01. 1.1.05.0001). 
V.V., N.Š., and D.A. acknowledge support from the Croatian Science Foundation project "Frequency comb cooling of atoms" (Grant No. IP-2018-01-9047).
\medskip

%


\bibliographystyle{MSP}

\clearpage
\includepdf[pages=-]{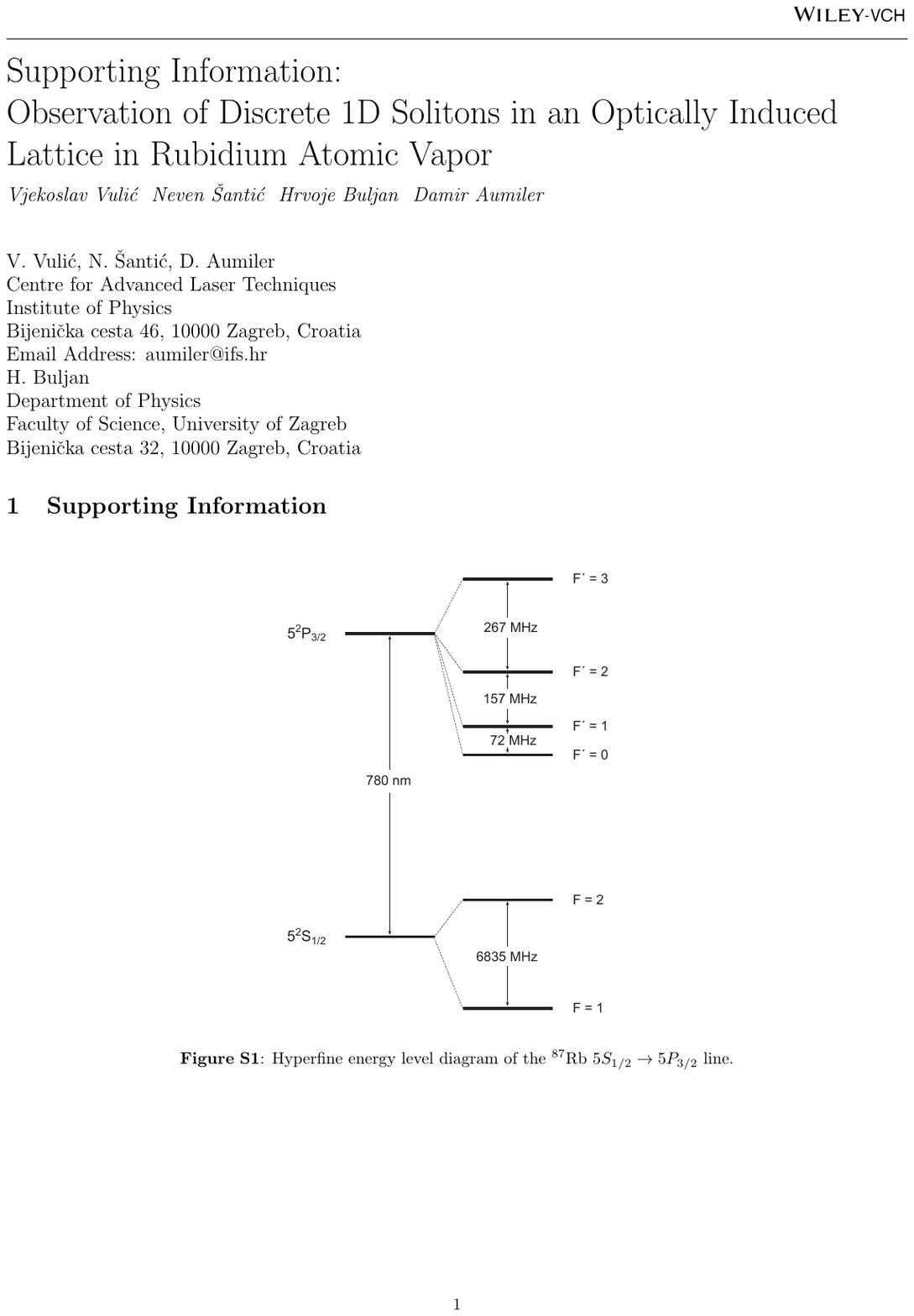}

\end{document}